\documentclass[useAMS,usenatbib]{mn2e}
\usepackage{amsmath}
\usepackage{mathtools}
\usepackage[utf8]{inputenc}
\usepackage{aas_macros}
\usepackage[toc,page]{appendix}
\usepackage{graphicx}

\newcommand{\ignore}[1]{}
\usepackage{color}
\long\def\symbolfootnote[#1]#2{\begingroup%
\def\thefootnote{\fnsymbol{footnote}}\footnote[#1]{#2}\endgroup}

\title[MDM]
  {Spectroscopic Follow-Up of the Hercules Aquila Cloud}
\author[I.T. Simion et al.]
  {Iulia T.~Simion$^{1,2}$\thanks{email: isimion@shao.ac.cn}, Vasily~Belokurov$^2$, Sergey E.~Koposov$^{3,2}$,   Allyson Sheffield$^4$,
  \newauthor
  Kathryn V.~Johnston$^5$\\
$^{1}$Key Laboratory for Research in Galaxies and Cosmology, Shanghai Astronomical Observatory, 80 Nandan Road, Shanghai 200030, \\ China\\
$^{2}$Institute of Astronomy, Madingley Road, Cambridge, CB3 0HA \\
$^{3}$Carnegie Mellon University, 5000 Forbes Avenue, Pittsburgh \\
$^{4}$Department of Natural Sciences, LaGuardia Community College, City University of New York, 31-10 Thomson Avenue, New York 10027 \\
$^{5}$Department of Astronomy, Columbia University, 550 West 120th Street, New York 10027
}
\date{accepted 19 February 2018;}

\pagerange{\pageref{firstpage}--\pageref{lastpage}} \pubyear{2017}

\def\LaTeX{L\kern-.36em\raise.3ex\hbox{a}\kern-.15em
    T\kern-.1667em\lower.7ex\hbox{E}\kern-.125emX}

\begin{document}

\label{firstpage}

\maketitle

\begin{abstract}
We designed a follow-up program to find the spectroscopic properties of the Hercules-Aquila Cloud (HAC) and test scenarios for its formation. We measured the radial velocities (RVs) of 45 RR Lyrae in the southern portion of the HAC using the facilities at the MDM observatory, producing the first large sample of velocities in the HAC. We found a double-peaked distribution in RVs, skewed slightly to negative velocities. We compared both the morphology of HAC projected onto the plane of the sky and the distribution of velocities in this structure outlined by RR Lyrae and other tracer populations at different distances to N-body simulations. We found that the behaviour is characteristic of an old, well-mixed accretion event with small apo-galactic radius. We cannot yet rule out other formation mechanisms for the HAC. However, if our interpretation is correct, HAC represents just a small portion of a much larger debris structure spread throughout the inner Galaxy whose distinct kinematic structure should be apparent in RV studies along many lines of sight.
\end{abstract}
\section{Introduction}
The halo of the Milky Way contains clear evidence of the merging processes that contributed to its formation: in the past 20 years, large numbers of stellar tracers from deep, wide-field imaging surveys have uncovered at least 20 streams and 6 cloud-like structures (for a list see \citealt{Be13} and \citealt{Grillmair2016}) lurking within a -previously thought- smooth Halo. These structures are predicted by models of hierarchical $\Lambda$CDM galaxy formation \citep[e.g.][]{Steinmetz2002, Bu05} which produce \textit{shells} when satellites are accreted on near-radial orbits \citep[see e.g.][]{He15, Pop2017} and \textit{streams} when satellites have more circular orbits \citep{Jo08}. Many efforts have been dedicated to the study of Galactic stellar streams but in recent years there has been an increased focus on the less studied and understood shells. While shells are frequently observed in external galaxies \citep[see e.g. NGC 747 and 7600 in][]{Tu99} only few candidates have been found in the Milky Way, which, when viewed from an internal perspective, take the shape of cloud-like, irregular, stellar overdensities. Candidate examples of these structure include the Triangulum-Andromeda stellar clouds, TriAnd1 and TriAnd2 \citep[e.g.][]{Sh14, Pr15}, the Virgo Overdensity \citep[e.g.][]{Ca12, Du14, Vivas2016}, the Pisces Overdensity \citep[e.g.][]{Wa09}, the Eridanus-Phoenix Overdensity \citep[][]{Li2016} and the Hercules-Aquila Cloud \citep[e.g.][]{Be07, Si14}.

The Hercules Aquila Cloud (HAC) was first discovered by \citet{Be07} as a  Main Sequence Turnoff  (MSTO) overdensity of stars in the Sloan Digital Sky Survey (SDSS) and it was described as a cloud because of its observed morphology. MSTO stars reveal a large structure beyond the Bulge, at line of sight distances larger than 10 kpc with a huge sky coverage between $20^{\circ}<l<70^{\circ}$ and $b$ up to $\pm 50^{\circ}$. However, the full extent of the Cloud was unknown: close to the Galactic plane ($|b| < 20^{\circ}$) the data is compromised by extinction and, in the Southern Hemisphere, only two SDSS DR5 stripes cross the field at two constant longitudes, l $= 31^{\circ}$ and $50^{\circ}$. The more contiguous SDSS sky coverage in the North (b$>20^{\circ}$), allowed \citet{Be07} to estimate the velocity of the Northern HAC ($v_{\mathrm{GSR}} \sim$ 180 km/s) and its metallicity, concluding the HAC was somewhat more metal rich than the M92 globular cluster ([Fe/H]$^{\mathrm{M92}}$ = -2.2 dex). 

RR Lyrae are reliable tracers of halo substructure thanks to their well defined Period-Luminosity metallicity relation which allows for accurate distance determinations, with less than 10\% uncertainty. Several studies using RR Lyrae \citep[][]{Wa09, Sesar2010, Si14, Iorio2017} from SDSS, Catalina Schmidt Survey (CSS) and Gaia+2MASS have helped build a map of the Cloud at intermediate latitudes, confirming it is asymmetric with respect to the Galactic plane: \citet{Si14} found a significant RRL overdensity in the Southern Cloud (8$\sigma$ level) at 17 kpc from the Sun, while in the Northern Hemisphere the structure was barely mappable ($3\sigma$ level significance) possibly due to the larger presence of interstellar dust (see the extinction contours in figure 3, \citealt{Si14}). The HAC RR Lyrae stars are mainly of type Oo I or fall in the Oosterhoff gap indicating the HAC progenitor resembles the main Halo population \citep[][]{Sesar2010, Si14}. An early metallicity measurement \citep[][]{Wa09} using RR Lyrae from Stripe 82, indicates the Southern HAC is slightly more metal rich than the Halo, with [Fe/H] $= -1.42 \pm 0.24$ dex; however, the velocity distribution of the MSTO stars in Stripe 82, is broad and mainly dominated by the thick disc contamination, at~100 km/s (\citealt{Wa09}, figure 17), not providing a clear measurement of the velocity distribution of this halo substructure.
\\
Before the distance to the Cloud was confirmed with RR Lyrae, \citet[][]{Larsen2008, Larsen2011} reported a structure between $\sim$ 1 -- 6 kpc from the Sun that was dubbed the “Hercules Thick Disk Cloud.” \citet{Larsen2011} suggested that the overdensity seen by \citet{Be07} was actually an artifact of the background-subtraction method and that the true stellar overdensity is actually part of the Disc. It has since been established the HAC resides in the Halo (see e.g. references in the previous paragraph) but its formation mechanism is not known. Recent studies with M giants have explored the possibility that some of the low-latitude Cloud-like structures (TriAnd1, TriAnd2 and Virgo Overdensity) with similar morphology to the HAC have originated in the Milky Way Disc and are made up by kicked-off stars from the Galactic Plane (\citealt{Johnston2017} and references therein). However, there have been no reports of M giant stars in the HAC, which, unlike the RR Lyrae, are abundant in the Galactic Disc. M giant stars are younger and more metal rich than the RR Lyrae and it is not surprising these different tracers do not trace the same structures (e.g. TriAnd contains M giants but not RR Lyrae). The strong presence of the typically old (older than 10 Gyrs) and metal poor RR Lyrae population indicates the HAC is probably related to an old accretion event instead, and does not originate from Disc stars. 

We have designed a spectroscopic follow-up program of the CSS RR Lyrae that lie in the peak of the Southern Hemisphere excess to investigate the kinematic signature of the Cloud. We show that in addition to a halo population modeled with a Gaussian, the observed velocity distribution requires two additional Gaussians; the peaks are situated at moderately large negative and positive radial velocities, and we interpret these components to be part of the Cloud. 

While the RR Lyrae are the ideal tracers for the Southern HAC because of their accurate distances and complete sky coverage, we also use Blue Horizontal Branch (BHBs) stars, Blue Stragglers (BS) and K giants from SDSS (which covers only a small area of the Southern HAC) to study their phase space distribution in the HAC.

The BHBs, as the RR Lyrae, are located on the Horizontal Branch in the Hertzsprung-Russell diagram; similarly, they are bright  ($M_{g} \sim$ +0.7 mag), old, metal-poor stars, therefore good tracers of halo substructure \citep[e.g.][]{Deason2014}. The HAC is revealed by the BHBs population in the Southern Galactic hemisphere section of the celestial equator (SDSS, Stripe 82) at (l,b)$\sim (50^{\circ}, -30^{\circ})$ and $\sim$ 18 kpc (figure 6, \citealt{Vickers2012}). The BS, on the other hand, are main-sequence stars with a wide range of absolute magnitudes that vary with metallicity, rendering their distance estimates much less accurate; even so, they have also been successfully used in Galactic halo studies \citep[e.g.][]{De11} and we decide to use them together with the BHBs to investigate the velocity signature of the Southern HAC. 

Unlike the aforementioned tracers, the intrinsic luminosities of K giants vary by two orders of magnitude, with colour and luminosity depending on stellar age and metallicity, leading to significant distance uncertainty. However, they have been used to quantify the level of substructure in the halo and the measurements are consistent with the BHBs results over the same ranges of distances \citep{Xu14}, leading us to believe that the phase space distribution of BHBs and K giants in the HAC should be comparable.

The paper is organised as follows: in Section 2 we describe the RR Lyrae selection criteria for spectroscopic follow-up,  data reduction,  velocity determination and the sample of SDSS tracers (BS, BHBs and K giants);  in Section 3 we discuss the multi-Gaussian decomposition of the velocity distribution and in Section 4 we interpret the results with the help of N-body simulations; finally, we summarise our conclusions in Section 5.

\section{Data}
The spectroscopic data originates from two sources, the CSS RR Lyrae follow-up program we have specifically designed to investigate the HAC and the SDSS/Sloan Extension for Galactic Understanding and Exploration (SEGUE) stellar spectra catalog.
\subsection{The RR Lyrae follow-up program}
\subsubsection{Targets selection: MDM}
The spectroscopic candidates were selected from the CSS catalogs (table~1 in \citealt{Drake2013a} and table 2 in \citealt{Drake2013b}) to lie in the peak of the HAC RRab excess in the Southern Hemisphere at  28$^{\circ}<l< 55^{\circ}$, -45$^{\circ}<b<-20^{\circ}$ and 15 $ <  D/$kpc $ <$ 20 \citep{Si14}. In total, we obtained 225 candidates, at least $\sim$34\% of which belong to the Cloud. The percentage was estimated comparing the number of observations to number of counts model predictions (see also figure 4, \citealt{Si14}). 

\subsubsection{Observations}
\begin{figure}
\hspace{-0.5cm}
\includegraphics[scale=0.29]{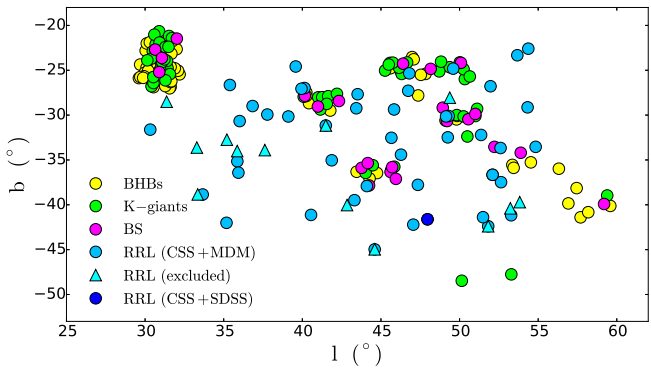}
\caption{Spectroscopic HAC candidates in Galactic coordinates. Observed RR Lyrae (final catalog targets: light blue circles; excluded targets, including three duplicate observations: cyan triangles) selected from the CSS survey with V $< $ 17.5 mag and 15 $ <  D/$kpc $ <$ 20. The deep blue circle is a CSS HAC candidate from SDSS. Blue Horizontal Branch (BHBs, yellow circles) stars have heliocentric within 12 $ < D/$kpc $ <$ 20, K-giants (green circles) within 12 $ < D/$kpc $ <$ 20 and Blue Stragglers (BS, magenta circles) within 10 $ <  D/$kpc $ <$ 20, are all selected from SDSS/SEGUE.}
\label{lb}
\end{figure}

The observations were taken with the Modular Spectrograph (ModSpec) on the 2.4\,m Hiltner telescope at the MDM observatory during six nights (labeled `n1' to `n6' in the tables of this manuscript) between the 29$^{th}$ of August 2014 and the 3$^{rd}$ of September 2014. Nights `n3' and `n4' were not fully used for observations because of technical problems and bad weather. In total, we took 54 observations marked in light blue circles (final catalog) and cyan triangles (unused targets) which we show in Galactic coordinates in Figure \ref{lb}.

The spectrograph was set-up with a 600 grooves/mm grating, central wavelength of 5300\AA, and wavelength coverage from 3800\AA~to 7300\AA~and resolution of R = $1700$. The detector was a 2048 $\times$ 2048 pixels CCD with 15 micron pixels. The wavelength coverage of the spectrograph allowed us to detect the $H_{\delta}$, $H_{\gamma}$,  $H_{\beta}$, $H_{\alpha}$ Balmer lines (4102\AA, 4341\AA , 4861\AA , 6563\AA ) and [OI] atmospheric emission lines (e.g., 5577\AA  ~and 6300\AA ), useful for checking for possible RV offsets. Given the resolution of the spectrograph and the dispersion of $\sim$ 2\AA/pixel the radial velocities could be measured accurately up to 10 km/s precision. 

The range of magnitudes required to probe heliocentric distances between 15 to 20 kpc with RR Lyrae is $16.5 < V_{0}/$ mag $< 17.5$ and to achieve a signal to noise of $\sim$ 20 per pixel with our instrument set-up an integration time of $\sim$ 750 seconds is needed.

To limit the integration time we only observed targets brighter than $V < 17.5$ mag, which meant observing preferentially closer targets. In practice, we used total integration times between 900 and 1800 seconds, depending on the seeing, airmass and target brightness.

To determine the wavelength solution, neon, xeon, argon and mercury comparison lamp spectra were taken throughout the night, specifically after long telescope slews to account for telescope flexure. In addition, a flat field (quartz lamp) and bias frames were obtained each night. 

\subsubsection{Data reduction}
The data reduction was performed using a collection of \textit{IRAF} tasks and \textit{Python} routines. 
Preprocessing of the spectra was carried out using the \textit{IRAF} \textit{ccdproc} task. Variations in the bias level along the CCD chip were removed using the overscan strip on a frame by frame basis. Biases were taken at the beginning and end of each night to verify that there are no significant drifts in the pedestal level. To correct for pixel-to-pixel sensitivity variations in the detector we divided each data frame by the normalised flat, which was created by fitting a low order polynomial to the median combined flats using the \textit{response} task. To increase the signal to noise of some exposures, three spectra were typically observed and co-added.
\begin{figure}
	\hspace{-0.4cm}
	 \includegraphics[scale = 0.29]{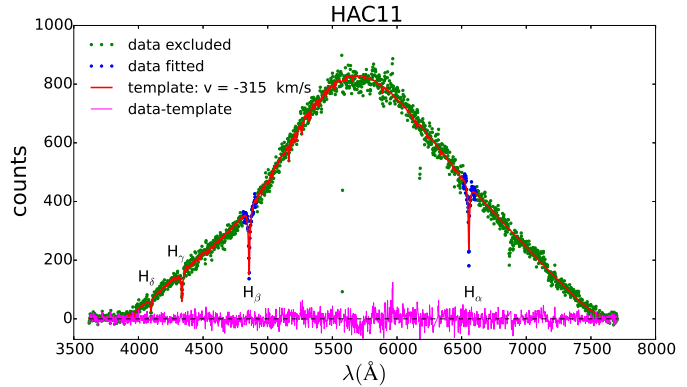} 
	\caption{Spectrum of target HAC11 (in green) and best fit template (in red). Four of the Balmer lines are identifiable in the spectrum but the velocity was determined from the fit of the $H_{\alpha}$ and $H_{\beta}$ lines only (shown in blue); the  $H_{\gamma}$ and  $H_{\delta}$ lines were not included in the fitting procedure as for several targets these lines cannot be used due to low signal-to-noise ratio in the data. In pink we show the residuals between the data and the template after excluding the [OI] atmospheric emission lines. } 
	\label{1dspec}
\end{figure}
The IRAF \textit{apall} and \textit{identify} tasks were used for one dimensional spectral extraction and wavelength calibration. We applied the dispersion solution using the \textit{dispcor} task and to correct for the Earth's motion with respect to the barycentre of the solar system we used the \textit{rvcorrect} task. The [OI] night sky emission lines were used to check for any systematic offsets in radial velocities for each night. We calculated the overall level of stability of our radial velocities ($\sim$10\,km/s on average) individually for each night (see Table \ref{errors}) and include it as systematic error in the total error budget of radial velocities given in Table~\ref{resultsMDM}.
\setcounter{table}{0}
\begin{table}
 \centering
 \begin{minipage}{80mm}
  \caption{Average variations in the overall level of stability of the velocities for each night, calculated using the [OI] night sky emission lines 5577\AA~and 6300\AA~. These systematic errors are part of the total errors budget of radial velocities given in Table~\ref{resultsMDM}. }
   \label{errors}
\begin{tabular}{l | c c c c c c }
 \hline
\setcounter{table}{0}
  observing night  &  n1 & n2 & n3 & n4 & n5 & n6 \\
   \hline
$\sigma_{obs}$ (km/s)  & 12.7 & 8.4 & 9.8 & 14.2 & 9.1 & 3.9 \\
\end{tabular}
\end{minipage}
\end{table}
\subsubsection{Velocity determination and spectral fitting}
To determine the heliocentric RV of each star we used a direct pixel-fitting method (e.g. \citealt{Ca04}; \citealt{Ko09}; \citealt{Ko11}) in which the spectrum was modeled with a template multiplying a normalising polynomial of degree $N$ (equation 2 in \citealt{Ko11}) and we found that $N = 15$ worked well for all targets. A sample of 90 templates were chosen from a library of synthetic spectra \citep{Mu05} with stellar atmospheric parameters that match the typical properties of RR Lyrae: $-2.5  <$ [Fe/H] $<$ -0.5; [$\alpha$/Fe] = 0.4;  5,500 $< T_{\mathrm{eff}}/$K $<$ 7,500; 2.0 $< \log(\textit{g}) < $ 4.0 \citet[][]{Smith2004}. 

We modeled two regions of the spectrum, one centred on $H_{\beta}$ (4811 - 4911 \AA)  and one on  $H_{\alpha}$ (6512 - 6612 \AA ), shown in blue in Figure \ref{1dspec} for the 'HAC11' target. The  $H_{\delta}$ and  $H_{\gamma}$ lines are discernible in this figure but for most stars they are too weak to be included in the fitting procedure.

We follow the method described in detail by \citet{Ko11}. The $\chi^{2}$ value was computed from a grid of templates and radial velocities $v$ between -600 and +400 km/s with a step of 1 km/s. The best fit template and RV are the ones that minimise $\chi^{2}$.  In column 7 of Table~\ref{resultsMDM} we report the velocities $v$ we obtained for each star. The uncertainty in the best fit $v$ was determined from the second derivative of the chi square function with respect to the RV near the minimum, $\sigma_{\mathrm{fit}}= 1/\sqrt{0.5 ~ d^{2} \chi^{2}/d v^{2}} $. 

\setcounter{table}{1}
\begin{table*}
 \centering
 \hspace*{-1.cm}
 \begin{minipage}{150mm}
   \caption[Properties of the Program stars]{Properties of the program stars observed at the MDM observatory. 225 HAC candidates were selected from CSS, using the following selection criterium: 28$^{\circ}$$<l<$ 55$^{\circ}$ , -45$^{\circ}$$< b <$-20$^{\circ}$ , V $< $17.5 mag, 15$ <  D$/kpc$ <$ 20. In total we made 57 observations and 3 targets were observed twice; 45 of the observations had phases $0.1 < \phi_{obs} < 0.85$ while 9 (maked with a * in the table and with cyan triangles in Figure \ref{lb}) fell outside this range and were discarded because their pulsation velocity corrections are uncertain. One of the 225 HAC candidates was observed by SDSS (deep blue circle in Figure \ref{lb}) and we include it in our sample. }
   \label{resultsMDM}
  \begin{tabular}{@{}llllllclccllll@{}}
 \hline
  night &  name &   RA &   Dec & D & $ \phi_{obs}$ & $v$ & $\chi_{red}^{2}$ & $v_{sys}$ & $v_{GSR}$ &$\sigma$ &  \\
    &   & $ (^{\circ}, J2000) $ & $ (^{\circ}, J2000) $ & (kpc) &  &  (km/s) &  & (km/s) & (km/s) & (km/s) &  \\
 \hline
n1 & HAC211 & 314.3614 & -7.0397 & 15.6 & 0.608 & -286.0 & 1.06 & -325.8 & -195.4 & 16.39 \\
n1 & HAC197 & 328.2793 & -5.4308 & 15.14 & 0.1048 & -173.0 & 1.33 & -155.5 & -25.0 & 15.11 \\
n1 & *HAC59 & 313.4314 & -12.531 & 16.07 & 0.0995 & -362.0 & 1.41 & -341.9 & -229.7 & 15.03 \\
n1 & HAC37 & 311.8052 & -11.0359 & 15.31 & 0.5106 & -336.0 & 1.55 & -375.7 & -258.5 & 14.85 \\
n1 & HAC152 & 322.1852 & 1.423 & 16.31 & 0.5095 & -181.0 & 1.51 & -213.7 & -59.4 & 15.36 \\
n1 & HAC22 & 310.0809 & -6.161 & 15.69 & 0.5052 & 99.0 & 1.25 & 62.3 & 195.2 & 15.43 \\
n1 & HAC27 & 310.6845 & -15.8723 & 15.99 & 0.1084 & -34.0 & 1.47 & -16.4 & 84.0 & 15.84 \\
n1 & HAC186 & 326.3078 & -8.3674 & 16.29 & 0.4982 & -26.0 & 1.24 & -61.8 & 60.5 & 15.31 \\
n1 & HAC91 & 315.9205 & -13.0186 & 15.84 & 0.5047 & -241.0 & 1.63 & -272.8 & -162.4 & 14.78 \\
n2 & HAC26 & 310.5762 & -9.6702 & 15.37 & 0.3661 & -364.0 & 1.25 & -377.4 & -255.8 & 11.61 \\
n2 & HAC131 & 319.4763 & -0.2823 & 17.11 & 0.3085 & -177.0 & 1.45 & -179.2 & -28.9 & 14.88 \\
n2 & *HAC210 & 313.5939 & -14.3483 & 15.53 & 0.0304 & 119.0 & 1.26 & 150.4 & 256.3 & 16.58 \\
n2 & HAC11 & 307.868 & -9.8056 & 15.66 & 0.1164 & -315.0 & 2.06 & -295.3 & -174.7 & 12.26 \\
n2 & HAC177 & 324.4688 & -2.2511 & 15.56 & 0.1032 & 45.0 & 1.81 & 63.2 & 205.7 & 12.51 \\
n2 & HAC9 & 307.7044 & -5.5085 & 15.34 & 0.5026 & 27.0 & 1.88 & -4.3 & 130.2 & 12.26 \\
n2 & HAC195 & 327.9145 & -11.2382 & 15.34 & 0.5046 & -329.0 & 2.97 & -356.1 & -244.3 & 12.63 \\
n2 & HAC96 & 316.7525 & -0.6645 & 15.23 & 0.1155 & -148.0 & 2.11 & -129.3 & 20.4 & 10.94 \\
n2 & HAC105 & 317.272 & -4.579 & 15.31 & 0.1139 & -273.0 & 2.05 & -256.4 & -118.5 & 11.23 \\
n2 & HAC102 & 317.1288 & -13.4841 & 15.31 & 0.4887 & 125.0 & 2.2 & 96.1 & 204.7 & 13.88 \\
n2 & HAC145 & 321.3978 & -8.286 & 15.55 & 0.5108 & 16.0 & 1.84 & -16.6 & 108.3 & 11.5 \\
n3 & HAC51 & 313.1055 & 6.2501 & 16.88 & 0.6 & -324.0 & 1.82 & -372.0 & -202.8 & 15.35 \\
n3 & HAC45 & 312.3731 & 2.3048 & 16.17 & 0.7213 & -163.0 & 1.77 & -222.8 & -64.3 & 20.83 \\
n3 & HAC52 & 313.176 & -1.1647 & 17.1 & 0.532 & -363.0 & 1.71 & -400.9 & -252.4 & 15.56 \\
n3 & HAC42 & 312.1074 & -3.8391 & 16.24 & 0.1131 & 13.0 & 1.47 & 30.4 & 170.8 & 13.55 \\
n3 & HAC156 & 322.5188 & -9.5651 & 16.21 & 0.5376 & 33.0 & 1.92 & -4.8 & 115.4 & 14.56 \\
n4 & *HAC204 & 308.0852 & -13.8239 & 17.74 & 0.9028 & 64.0 & 1.47 & 5.4 & 112.5 & 23.48 \\
n4 & HAC94 & 316.7138 & -0.8093 & 15.71 & 0.6281 & -263.0 & 1.69 & -315.3 & -166.0 & 16.5 \\
n4 & HAC191 & 327.2894 & -5.0996 & 18.04 & 0.5205 & -116.0 & 1.56 & -154.1 & -21.9 & 19.27 \\
n4 & HAC194 & 327.836 & -3.8522 & 17.64 & 0.5425 & -372.0 & 1.54 & -406.5 & -270.9 & 18.33 \\
n4 & *HAC218 & 318.6157 & -16.3178 & 15.2 & 0.0778 & 82.0 & 1.23 & 101.8 & 200.2 & 16.12 \\
n4 & HAC21 & 310.0706 & -6.3517 & 15.29 & 0.5129 & -84.0 & 1.48 & -111.6 & 20.7 & 15.23 \\
n4 & HAC208 & 311.8013 & -9.3492 & 18.12 & 0.6435 & 65.0 & 1.46 & 10.5 & 133.3 & 17.43 \\
n4 & HAC166 & 323.0617 & -12.1824 & 16.58 & 0.5222 & -23.0 & 1.57 & -51.0 & 60.2 & 17.26 \\
n4 & HAC154 & 322.215 & -16.2156 & 16.33 & 0.5205 & -372.0 & 1.77 & -402.5 & -304.8 & 15.86 \\
n4 & HAC120 & 318.7191 & -16.0923 & 15.87 & 0.5166 & 30.0 & 1.76 & 2.3 & 101.5 & 16.52 \\
n4 & HAC121 & 318.7578 & -1.944 & 16.49 & 0.1209 & -216.0 & 1.9 & -198.5 & -53.0 & 15.71 \\
n4 & HAC187 & 326.8427 & -2.699 & 16.34 & 0.8679 & 82.0 & 1.51 & 19.5 & 159.3 & 19.03 \\
n4 & *HAC189 & 327.1678 & -3.4616 & 15.36 & 0.0652 & -481.0 & 1.68 & -459.7 & -322.4 & 15.33 \\
n4 & HAC169 & 323.5458 & -2.1666 & 17.64 & 0.4298 & -123.0 & 1.46 & -145.0 & -1.8 & 17.54 \\
n5 & HAC31 & 311.5271 & -0.1341 & 18.22 & 0.5134 & -375.0 & 1.31 & -412.5 & -261.0 & 13.71 \\
n5 & HAC48 & 312.8408 & 7.1898 & 17.84 & 0.4959 & 61.0 & 1.38 & 29.0 & 200.6 & 13.66 \\
n5 & HAC115 & 318.2882 & 3.5305 & 17.02 & 0.5272 & -169.0 & 1.75 & -203.5 & -42.1 & 13.83 \\
n5 & HAC60 & 313.4504 & -4.6574 & 16.21 & 0.8426 & -160.0 & 1.35 & -234.3 & -96.4 & 16.46 \\
n5 & HAC158 & 322.5319 & -5.9797 & 17.27 & 0.6726 & -285.0 & 1.59 & -332.5 & -200.6 & 13.3 \\
n5 & HAC113 & 317.9846 & -8.5446 & 17.6 & 0.673 & -127.0 & 3.13 & -185.9 & -60.8 & 18.85 \\
n5 & HAC71 & 314.591 & -2.8935 & 17.4 & 0.4947 & -393.0 & 1.57 & -424.3 & -281.0 & 12.05 \\
n5 & *HAC76 & 314.8484 & -12.5544 & 17.54 & 0.9257 & 77.0 & 2.11 & 10.9 & 123.0 & 16.6 \\
n5 & *HAC87 & 315.3761 & -11.1874 & 16.91 & 0.0678 & -102.0 & 1.28 & -77.9 & 38.8 & 14.69 \\
n5 & HAC47 & 312.5178 & -8.409 & 17.5 & 0.6665 & -97.0 & 2.1 & -150.9 & -25.0 & 12.25 \\
n5 & *HAC163 & 322.8444 & -10.1355 & 17.47 & 0.0782 & -193.0 & 2.2 & -167.8 & -49.5 & 10.62 \\
n5 & HAC126 & 319.1649 & -5.063 & 16.99 & 0.5857 & -197.0 & 1.57 & -240.2 & -104.3 & 14.9 \\
n5 & *HAC80 & 315.0363 & 0.4506 & 17.32 & 0.0093 & -265.0 & 1.88 & -285.6 & -132.4 & 11.24 \\
n5 & HAC169 & 323.5458 & -2.1666 & 17.64 & 0.5426 & -71.0 & 1.28 & -113.0 & 30.2 & 19.46 \\
n5 & HAC82 & 315.1755 & 3.0767 & 17.02 & 0.4777 & -62.0 & 1.47 & -88.8 & 71.9 & 14.44 \\
n6 & HAC143 & 321.2616 & -0.1949 & 17.19 & 0.5346 & -65.0 & 1.75 & -94.8 & 55.2 & 14.55 \\
SDSS & HAC184 & 326.1226 & -7.4878 & 16.19 & - & -& - & - & 5.3 & 14.3 \\
\hline
    \end{tabular}
 \end{minipage}
\end{table*}

\subsubsection{Systemic velocities}
RR Lyrae are pulsating variable stars and to find their systemic velocity $v_{\mathrm{sys}}$ (the center-of-mass RV), we need to subtract the velocity due to envelope pulsations 
\begin{equation}
v_{\mathrm{puls} }= \frac{A_{rv}^{\alpha}T_{\alpha}(\phi_{obs}) + A_{rv}^{\beta}T_{\beta}(\phi_{\mathrm{obs}})}{2} 
\label{eq1}
\end{equation} 
from the measured velocity $v$. $A_{rv}$ is the amplitude of the radial velocity curve, $T(\phi_{\mathrm{obs}})$ is the template radial velocity curve which describes the changes in RV as a function of pulsation phase and $\phi_{\mathrm{obs}}$ is the phase at the time of the observation. This method  was successfully employed in recent years by e.g. \citet{Drake2013a}, \citet{Vivas2016} and \citet{Ablimit2017} on SDSS, SOAR, WIYN and LAMOST spectra and is described into detail by \citet{Se12}.

The phase of each star at the time of the observation is given by
\begin{equation}
\phi_{\mathrm{obs}} = \frac{\mod((\mathrm{MJD}_{\mathrm{obs}} - \eta),P)}{P}
\end{equation}
where $\eta$ is the ephemeris (the MJD of maximum brightness), $P$ is the period of the light curve and MJD$_{\mathrm{obs}}$ is the Modified Julian Date at the time of the observation \citep{Drake2013a}. The RVs vary rapidly at low and high $\phi$, so we remove targets with phases $\phi_{\mathrm{obs}}<0.1$ or $\phi_{\mathrm{obs}}>0.85$ (marked by * in Tables~\ref{resultsMDM} and shown with cyan triangles in Figure \ref{lb}) from further analysis. Two of the stars with 'extreme' phases, HAC211 and HAC197, were observed twice and we include in the table only the measurement with  $0.1<\phi_{\mathrm{obs}}<0.85$. The only target observed twice at borderline 'acceptable' phases is HAC195 and the velocity measurements from the two nights, $v_{\mathrm{GSR}} = -253.2 \pm 15 $ km/s (n1, $\phi_{obs} = 0.09$)  and $v_{\mathrm{GSR}} \ = -244 \pm 13 $ km/s (n2, $\phi_{\mathrm{obs}} = 0.50$), agree within one sigma. 

Knowing the phase at which we observed the RRL with ModSpec, we then calculate the values of the template RV curves $T_{\alpha}(\phi_{\mathrm{obs}})$ and $T_{\beta}(\phi_{\mathrm{obs}})$ obtained from measurements of $H_{\alpha}$ and $H_{\beta}$ lines, assuming that at phase of 0.27 the RV is equal to the systemic velocity (see \citealt{sheffield18}). $A^{\alpha}_{rv}$ and  $A^{\beta}_{rv}$, are obtained from equations 3 and 4 in \citet{Se12} which provide their relation to the V-band light curves amplitudes $A_{V}$. Because the Balmer lines form at different heights in the stellar atmosphere and have different velocities as a function of pulsation phase, the most precise systemic velocity should be obtained by averaging the systemic velocities that are measured from individual lines \citep{Se12, Sesar2013}. However, because our stars are faint, we simultaneously fit the $H_{\alpha}$ and $H_{\beta}$ lines with the template spectrum to obtain $v$, from which we then subtract the velocity due to pulsations $v_{\mathrm{puls}}$ (Equation \ref{eq1}) calculated from two Balmer lines. 
\citet{Se12} estimate the level of uncertainty introduced by using more than one Balmer line in a cross-correlation and their figure 3  shows the standard deviation of radial velocities of various Balmer lines as a function of phase for RR Lyrae stars with different V-band amplitudes (the solid line shows the uncertainty when the $H_{\alpha}$ and $H_{\beta}$ lines
are used). In general, the uncertainty in velocity introduced by using more than one Balmer line is lowest for phases earlier
than 0.6, and it decreases with increasing V-band amplitude; from figure 3 \citep{Se12}, we roughly approximate it with: $\sigma_{\mathrm{puls}} = 2$ km/s for $\phi_{\mathrm{obs}}<0.4$ ,  $\sigma_{\mathrm{puls}} = 0$ km/s for $\phi_{\mathrm{obs}}\approx 0.5$ and a linear increase from $\sigma_{\mathrm{puls}}=$0 to 14 km/s for $0.5<\phi_{\mathrm{obs}}<0.95$. The final error reported in last column of Table \ref{resultsMDM} is $\sigma = \sqrt{\sigma_{\mathrm{obs}}^{2} + \sigma_{\mathrm{fit}}^{2}+\sigma_{\mathrm{puls}}^{2}}$. 

\subsection{SDSS/SEGUE tracers selection} 
\subsubsection{SDSS RR Lyrae}
\citet{Drake2013a} published a catalog of CSS RR Lyrae with SDSS spectra. One of these stars is within our selection criteria for the HAC candidates and we include it in the RRL sample (it is listed on last line of Table \ref{resultsMDM} and marked in dark blue in Figure \ref{lb}). 

\subsubsection{Blue Horizontal Branch Stars}
We select a subsample of BHBs (yellow circles in Figure \ref{lb}) from the catalog provided by \citet{Xu11}, with $12 < D$/kpc $<20$, $25^{\circ}<l<60^{\circ}$,  $-50^{\circ}<b<-20^{\circ}$, at distances $Z<-5$ kpc below the Galactic Plane to minimise Disc contamination, and obtain N = 70 targets. We relax the distance range used for RR Lyrae because the full extent of the HAC debris is not well constrained and, as mentioned in the introduction, different stellar populations do not always trace the same structures. In addition, the SDSS covers a smaller section of the HAC in comparison with our follow up survey (Figure \ref{lb}).
\subsubsection{Blue Stragglers}
\begin{figure*}
\hspace{-0.5cm}
\includegraphics[scale=0.25]{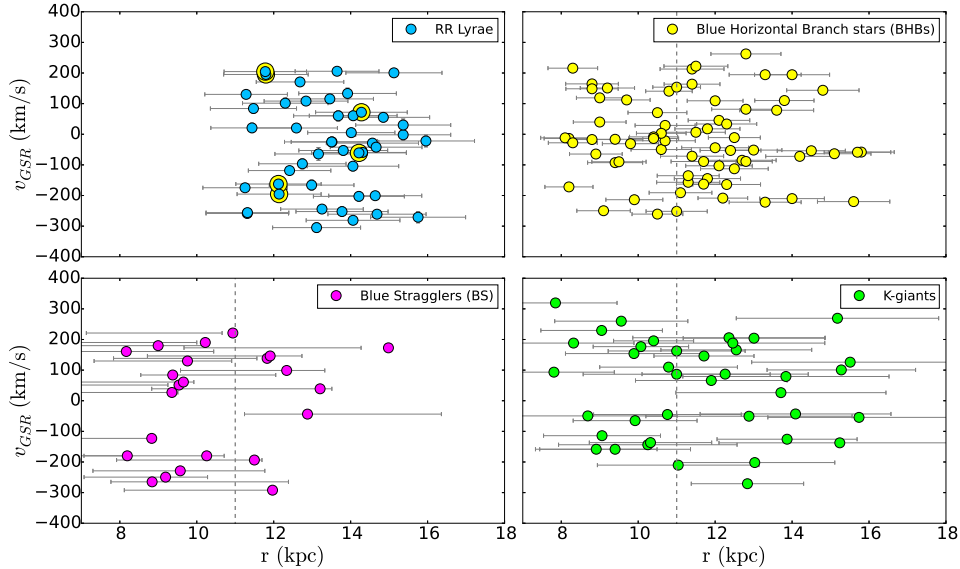}
\caption{Radial phase-space diagram with error bars for RR Lyrae (top left panel), BHBs (top right), BS (bottom left)  and K-giants  (bottom right). The large error bars on the galactocentric distance limits the spatial information we can extract from the samples. The yellow circles in the top left panel mark the 6 RR Lyrae of Oo II type. The distance uncertainties for RR Lyrae are $\sim$ 7\% while for BHBs are $\sim$5\% \citep{Xu11}. The dotted line at $r = 11$ kpc marks the galactocentric distance below which we do not have RR Lyrae in the spectroscopic sample. The bars on the BS markers indicate the distance the BS would have with a metallicity of -1 (upper limit) or -2.5 dex (lower limit). The distance uncertainties for K-giants are provided by \citet{Xu14}.}
\label{vels}
\end{figure*}

We select BS from SDSS/SEGUE (DR9) by their temperature, 7500 $<T_{\mathrm{eff}}$/K $< 9300$, and surface gravity, $4.0 < \log(g) < 4.6$  (see Figure 2 in \citealt{Be13}). \citet{Ki94} derived an absolute magnitude-colour-metallicity relation which we used to estimate $M_{g}$ by adopting the SEGUE metallicity values for stars with $[Fe/H] < -0.8$ dex  where the transformations are valid. We obtained $M_{g} \approx 2.7$ mag, $\sim$2 magnitudes fainter than the BHBs, signifying that the BSs are poor tracers of halo substructure at large distances; moreover, the magnitude-metallicity dependence introduces high uncertainties in the distance determination as shown by the error bars in the bottom left panel of Figure~\ref{vels}. The upper/lower limit of the error bars in the figure indicate the distance of the BS at a metallicity of -1/-2.5 dex . 

In the Heliocentric distance range $15<D$/kpc$<20$, where we found the RR Lyrae excess, there are only 5 spectroscopically observed BSs. However, due to significant errors in BS distances and limited SDSS sky coverage, we loosen the selection criteria to a wider range of Heliocentric distances $10<D$/kpc $ <20$, within $25^{\circ}<l<60^{\circ}$, $-50^{\circ}<b<-20^{\circ}$ and $Z < -5$ kpc. 

\subsubsection{K giants}
\citet{Xu14} compiled a catalog of 6036 K giants with stellar atmospheric parameters and distance determinations, most of which are members of the Milky Way's stellar halo. We chose the HAC candidates (green circles in Figure \ref{lb}) to lie at distances between $12<D/$kpc$<20$. To minimise contamination from thick disc dwarfs in our sample we only selected stars with a probability (provided by \citealt{Xu14}) higher than 80\% of clearly being red-giant branch stars and $Z<-5$ kpc. 
\begin{figure}
\includegraphics[scale=0.7]{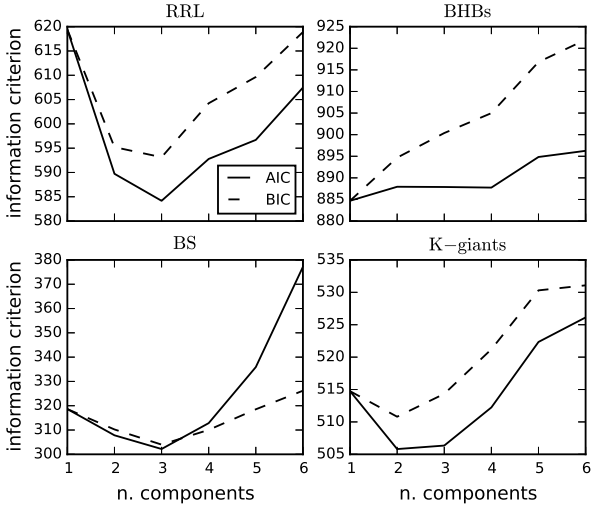} 
\caption[Best fit model for the Galactocentric radial velocities distribution of K giants]{ Model selection criteria AIC and BIC as a function of the number of components. They are minimised for 3 components (RR Lyrae), 1 component (BHBs), 2 components (K-giants) and 3 components (BS).}
\label{all_AIC}
\end{figure}
\begin{figure*}
\hspace{-1.1cm}
\includegraphics[scale=0.23]{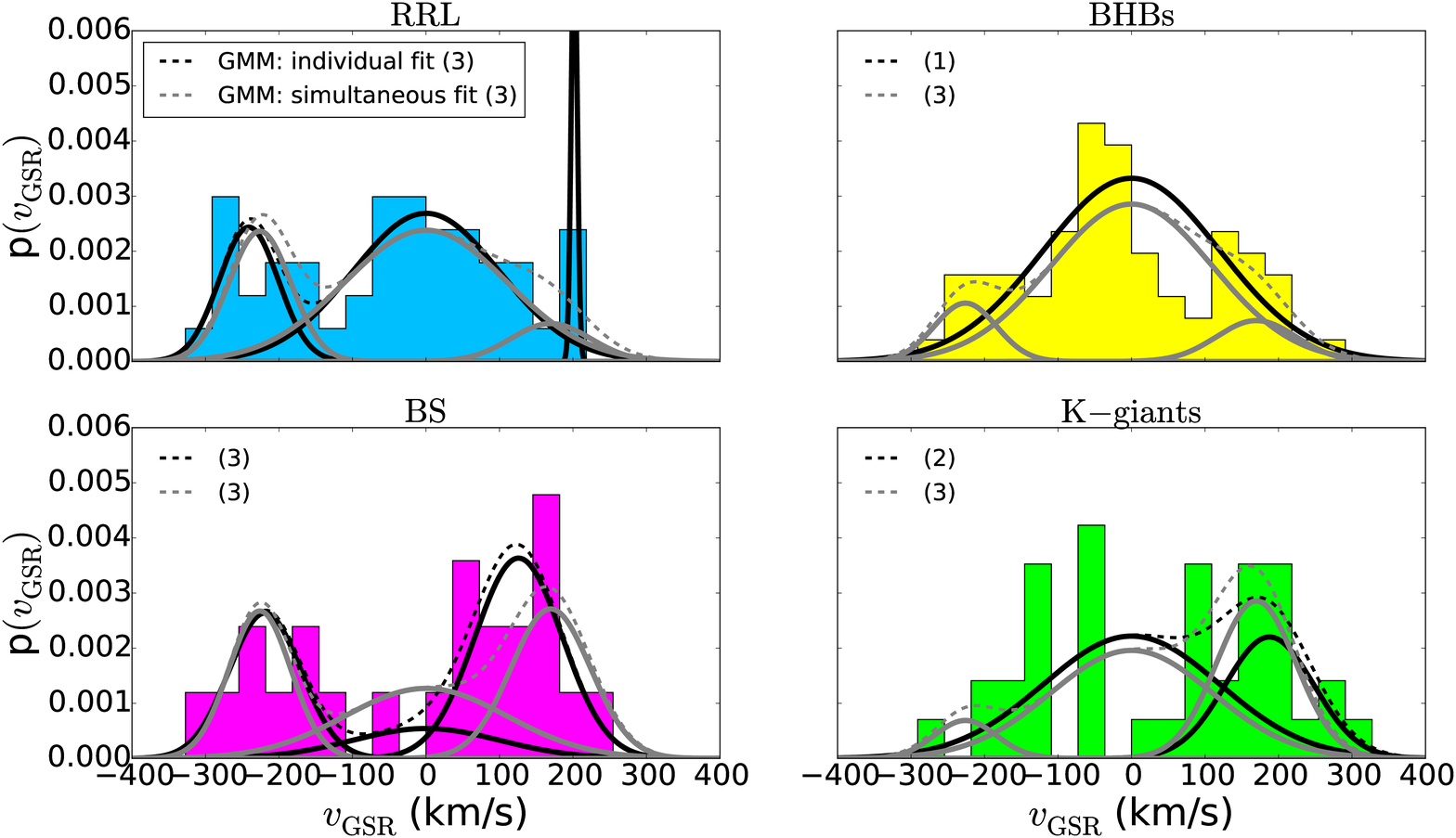} 
\hspace{-0.2cm}
\includegraphics[scale=0.32]{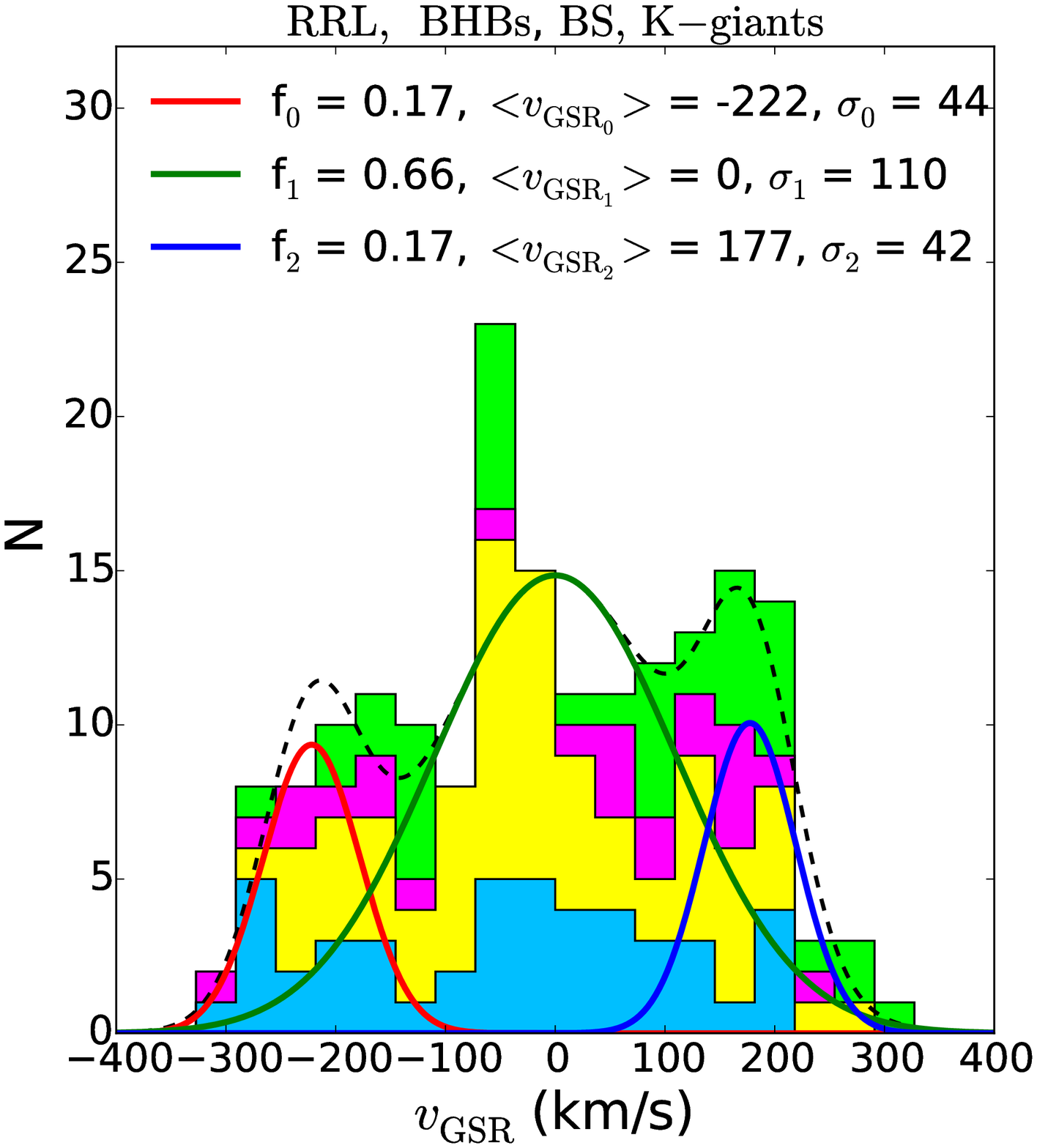}
\caption{4 left panels: RV distribution for the RR Lyrae (top left panel), BHBs (top right), BS (bottom left) and K-giants (bottom right) shown in Figures \ref{lb} and \ref{vels}. The best fit GMM is marked with dotted lines and each Gaussian in the mixture is marked with continuous lines; the number of Gaussians is specified in the legend. We fit the RV distribution of each tracer individually (black lines) and list the results in first 4 rows of Table \ref{EMtable}. The gray lines (dashed and continuous) refer to the simultaneous fit of all four tracer populations (fifth row in Table \ref{EMtable}) with 3 Gaussians, where each one has a different normalisation $f$. Right panel: RV distribution of all 4 populations combined and the best fit GMM which has 3 normalisations (given in the legend), as opposed to the GMM in the left panels which has 12  (listed in Table \ref{EMtable}).
}
\label{GM}
\end{figure*}

\section{Results: velocity distribution of the HAC}
In this section we explore the HAC kinematic signature revealed by the tracers selected in Section 2. The RR Lyrae and BS galactocentric velocities ($v_{\mathrm{GSR}}$) were converted from heliocentric velocities using equation 5 in \citet{Xu08} for consistency with the \citet{Xu11, Xu14} catalogs which use the same formula. The HAC does not immediately stand out from the halo field stars as an easily identifiable feature in velocity space, so we perform a Gaussian Mixture Model (GMM) decomposition of the velocity distribution of the halo tracers used.

 The likelihood of a velocity $v_{\mathrm{GSR}}^{i}$ for a GMM is given by 
 \begin{equation}
 p(v_{\mathrm{GSR}}^{i}|\theta) = \sum_{j=1}^{M} f_{j} N({v_{GSR}|\mu_{j}, \sigma_{j}})
 \end{equation}
where $\theta$ is the vector of parameters that needs to be estimated for the whole data set and includes the normalisation factors for each Gaussian $f_{j}$ and its Gaussian means and standard deviations $\mu_{j}$ and $\sigma_{j}$. This  likelihood is maximised using the expectation maximisation algorithm (EM) by \citet{Dempster77}. 

The model with the maximum likelihood ln$L_{\mathrm{max}}$, provides the best description of the data. However, the number of parameters increases as we add Gaussians to the mixture, and in order to choose the best model overall, we need a model comparison technique that ‘penalises’ models with too many parameters. To find the optimal number of Gaussians that best describe the observed velocity distribution, we evaluate the Akaike information criterion (AIC) and the Bayesian information criterion (BIC) for different values of M; we choose the model where the AIC or BIC is smallest (see top left panel in Figure \ref{all_AIC}, which shows the AIC and BIC as a function of M). The AIC and BIC are two analogues methods for comparing models, and they depend on the number of model parameters $k$, data points $N$ and the maximum value of the data likelihood: AIC$ = -$2ln$L_{max}$ + 2$k$, BIC = -2ln$L_{\mathrm{max}}$ + $k$ln($N$).

Studies with BHBs, BS and RR Lyrae \citep[e.g.][]{Xu08, Br10, Drake2013a} find that the field stellar halo population has a Gaussian velocity distribution roughly centred on 0 km/s with a velocity dispersion in the range 100 and 120 km/s, at Galactocentric radii $r < 20 $ kpc. According to equation 6 in \citet{Br10}, the value is close to $\sigma_{v}=103$km/s at r $=$ 16 kpc. We therefore include in our GMM a Gaussian with fixed mean at 0 km/s and a fixed velocity dispersion, as explained below.

\subsection{RR Lyrae}
In the top left panel of Figure \ref{vels} we show the radial velocities as a function of galacto-centric distance $r$, where the horizontal bars are the distance uncertainties. In this phase-space configuration, debris from accreted satellites is usually distributed in a typical 'bell-shape' (e.g. figure 2 in \citealt{Pop2017}) and is expected to overlap with the in situ halo population - given our low number of tracers it is probably not surprising no well-defined substructure pops out immediately. We refer the reader to the Discussion section for an interpretation of this figure. 
In the same plot, the yellow circles identify the 6 Oosterhoff type II RR Lyrae observed, which amount to only 11\% of the spectroscopic sample; the remaining 89\% targets are of Oosterhoff type I, in agreement with the conclusions drawn by \citet{Si14} who found that the bulk of RRL in the Cloud are type I, the dominant type in the Galactic halo ($\sim$75\%) .

The $v_{\mathrm{GSR}}$ distribution (blue histogram in Figure \ref{GM}) is best described by a GM with $M=3$ Gaussians (black solid curve overlaid on the histogram) where one represents the smooth halo population and the other two are structures centred at $\mu = -241$ km/s and $\mu = 202$ km/s. The central Gaussian is centred on $\mu = 0$ km/s and $\sigma$ is selected from a range of values between 100 and 120 km/s, with a step of 5 km/s, such that the AIC is minimal. The optimal number of Gaussians $M$ was found computing the AIC as a function of M (see Figure \ref{all_AIC}). The properties of the three velocity components are given on the top row of Table~\ref{EMtable}. We note that only 4 targets have high probability of belonging to the third population.  

\begin{table*}
 \centering
 \hspace*{-1.cm}
 \begin{minipage}{160mm}
   \caption[Expectation Maximisation parameters]{The maximum likelihood parameters for the Gaussian mixtures model for single population fits and 3 and 4 populations simultaneous fits. M is the number of Gaussians and N is the number of stars for each tracer. All tracers were selected to be at a distance of at least 5 kpc below the Galactic plane, $Z < -5$ kpc, and their coordinates are shown in Figure \ref{lb}. In the simultaneous fit the centres and the widths of the Gaussians are the same for all populations while the normalisations are fitted independently and are labeled $ f_{\mathrm{RR}}$, $f_{\mathrm{BHBs}}$, $f_{\mathrm{K}}$, $f_{\mathrm{BS}}$.}
    \label{EMtable}
         \begin{tabular}{@{}lllccccc@{}}
 \hline
  tracer & spectroscopic & selection & N & M & $f$ &  $<v_{\mathrm{GSR}}>$ & $\sigma$   \\
 & survey & HAC candidates &  &  &  &   (km/s) &  (km/s)  \\
 \hline
RR Lyrae & MDM, SEGUE & $15<D$/kpc$<20$ & 46 & 3 & 0.25 & -241 & 40  \\
 & & & &  & 0.08 & 202 & 4  \\
 & & & &  & 0.67 & 0 & 100  \\
  \hline
 K giants & SEGUE (DR9) &  $12<D$/kpc$<20$ & 39 & 2 & 0.33 & 188 & 60  \\
 & &P $> $0.80 &  & & 0.67 & 0 & 120  \\
 \hline
BHBs & SEGUE (DR8) &  $12<D$/kpc$<20$ & 70 & 1 & 1.00 & 0 & 120  \\
 \hline
BS &  SEGUE (DR9) &$10<D$/kpc$<20$ & 23 & 3 & 0.32 & -219 & 49  \\
 & &  $[Fe/H] < -0.8$ dex &  &  & 0.54 & 126 & 59   \\
  & &  &  &  & 0.14 & 0 & 100   \\
   \hline
 4 populations & &  & & 3 &$ f_{\mathrm{RR}}$, $f_{\mathrm{BHBs}}$, $f_{\mathrm{K}}$ , $f_{\mathrm{BS}}$ & & \\
 & & & &  & 0.25, 0.11, 0.07, 0.28 & -226 & 42  \\
 & & & &  &  0.66, 0.79, 0.54,0.35 & 0 & 110  \\
 & & & &  & 0.09, 0.10, 0.39,0.37 & 170 & 54  \\
 \hline
 3 populations: & & $15<D$/kpc$<20$ &  & 3 &$ f_{RR}$, $f_{\mathrm{BHBs}}$, $f_{\mathrm{K}}$ & & \\
- RR Lyrae & & & 46 &  & 0.23, 0.08, 0.00 & -236 & 39  \\
- BHBs & & & 38 &   & 0.69, 0.86, 0.93 & 0 & 115 \\
- K-giants & & & 18 &  & 0.08, 0.05, 0.07 & 200 & 6  \\
  \hline
    \end{tabular}
 \end{minipage}
\end{table*}
\subsection{Blue Horizontal Branch stars}
In the upper right panel of Figure \ref{vels} we show the sample of BHBs in a radial phase-space diagram, with distance uncertainties of 5\%. In the figure we mark the galactocentric distance $r = 11$ kpc (dotted line) below which we did not obtain RR Lyrae spectra. Their galactocentric velocity distribution (yellow histogram in Figure \ref{GM}) is consistent with a single Gaussian with $\sigma = 120$ km/s, centred on $\mu = 0$ km/s (see the model selection criteria in the top right panel of Figure \ref{all_AIC}). \citealt{Xu11} found that BHBs do not reveal much substructure for galactocentric distances $r_{\mathrm{gc}} < 20$ kpc so our result showing that the BHBs distribution can be described with a single Gaussian with $\sigma$ typical of the smooth halo population, is not surprising.

\subsection{Blue stragglers}
In the radial phase-space diagram of BS (bottom left panel of Figure \ref{vels}), there are two groups of stars at positive and negative velocities but their distances, $D < 14$ kpc, are closer than the HAC distance. The large distance uncertainties show we do not expect a systematic offset due to the absolute magnitude - metallicity dependence. The best GM model requires 3 Gaussians (AIC/BIC values are shown in the bottom right panel of Figure \ref{all_AIC}) with one component at negative velocities, $\mu = -219$ km/s and one at positive velocities, $\mu=126$ km/s.

\subsection{K-giants}
The galactocentric velocity distribution of the 39 selected HAC K-giants candidates as a function of distance $r$ is shown in the bottom right panel of Figure~\ref{vels}.
\begin{figure}
\centering
\includegraphics[scale=1.3]{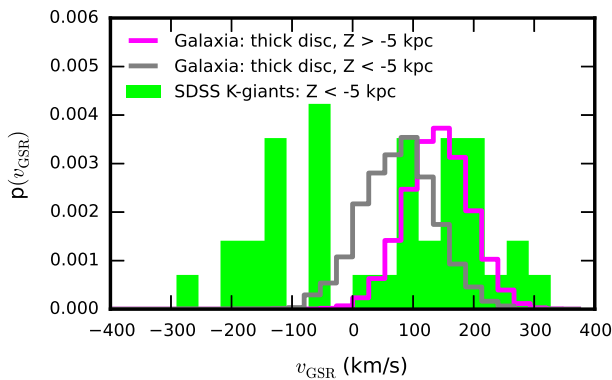}
\caption{RV distribution for the selected sample of K-giants (green). For reference, we show the velocity distribution of thick disc stars simulated with $Galaxia$, within the same distances and Galactic coordinates as our sample ( $Z<-5$ kpc, gray histogram) but also for stars closer to the Galactic plane, with $Z>-5$ kpc (magenta). The distribution close to the Galactic plane has a mean velocity of 167 km/s, while the one further from the plane, of 103 km/s.}
\label{Kdisc}
\end{figure}

The best GM model requires 2 Gaussians (bottom right panel of Figure \ref{all_AIC}) to fit the velocity distribution, where one represents the halo population and the  other  one is centred at positive velocities (see Figure \ref{GM} and best fit results in Table \ref{EMtable}). To minimise contamination from thick disc dwarfs in our sample we only chose stars with a probability higher than 80\% of being  K giants from the catalog provided by \citet{Xu14}. 

We take into account the possibility that the latter population is the thick disc; the disc at $l = 40^{\circ}$, $b = -30^{\circ}$ would have a velocity of $\sim$100 km/s ($v_{\mathrm{disc}} =180\sin(l)\cos(b)$); a test with $Galaxia$ \citep{Sh11} which can easily generate thick disc particles in the same region, indicates their velocity distribution peaks at 103 km/s for $Z < -5$ kpc (gray histogram in Figure \ref{Kdisc}) and 167 km/s for $Z > -5$ kpc (magenta histogram). It is therefore possible that the high velocity peak is partly due to thick disc stars contamination from the $Z > -5$ kpc region.
\subsection{Simultaneous fit of multiple populations}
We make the assumption that a Gaussian mixture with three components can describe the velocity distribution of all tracers and perform a simultaneous fit for all 4 populations considered (see the gray curves in the 4 left panels of Figure \ref{GM}), fitting the normalisations of the three Gaussians independently for each population but the centres and the dispersions simultaneously for the three Gaussians in the mixture. The results are reported in Table \ref{EMtable}: in addition to the Halo population best described with a Gaussian with width $\sigma_{\mathrm{halo}} = 110$ km/s, all tracers allow for two extra components at negative ($\mu = -226$km/s) and positive ($\mu=170$km/s) velocities respectively. The AIC of the simultaneous fit  (AIC = 2250) is comparable to the sum of the AIC of the 4 individual fits (AIC = 2270) an indication that our assumption is correct (note that we have a smaller number of free parameters in the simultaneous fit which decreases the value of the AIC). For completeness, we also show the velocity distribution of all 4 populations combined in one single plot (right panel of Figure \ref{GM}) and, overlaid, the result of a 3 Gaussians Mixture fit (dotted curve) to the total RV distribution. As opposed to the previous GMM which contained 12 normalisations $f_{i}$, this model only has 3, [$f_{0}$, $f_{1}$, $f_{2}$]$=$[0.17,0.66,0.17], where the biggest contribution is made by the halo population at 66\%. As expected, the velocity dispersions and mean velocities are similar in the two models (see legend in the right panel of Figure \ref{GM} and Table \ref{EMtable}). 

We also perform a simultaneous fit with 3 Gaussians for the RRL, K-giant and BHB populations alone (excluding the BS stars) and report the values in Table~\ref{EMtable}. Here we applied the same heliocentric distance cut on all populations, $15<D$/kpc$<20$ (in this region the BS population is too scarce to be included in the fit), to ensure that all the HAC candidates are within the same region, even though their sky coverage is different (RRL follow the CSS footprint while the rest of the tracers the SEGUE footprint). We again fit simultaneously the centres and widths of the Gaussians while the normalisations are fitted independently and we report the values on the last row of Table \ref{EMtable}. The best fit model required a (fixed) $\sigma_{halo} = 115 $ km/s (but the AIC values were almost identical for smaller $\sigma$) and no cold component for the K-giant population. \\
In all the models presented above (and listed in Table \ref{EMtable}), we can expect the central velocity peak to be caused by the field stellar halo stars and the positive and the negative velocity peaks to be linked to the HAC, in absence of other known structures in the region. The Sagittarius stream, with its almost ubiquitous coverage of the sky, could also contribute minimally to the velocity distribution: according to the  \citet{Law2010} model, a very small fraction of the stream stars ($\sim$1\%) are indeed situated at $D<20$ kpc in the region of the sky we have considered. These stars have a mean velocity of 240 km/s so they would (minimally) contribute to the high velocity peak. In addition, Disc stars could also be a contaminant at positive velocities (Figure~\ref{Kdisc}) especially for the least reliable tracers (e.g. K-giants). \\

\section{Discussion}
\begin{figure*}
\hspace{-0.8cm}
\includegraphics[scale=0.8]{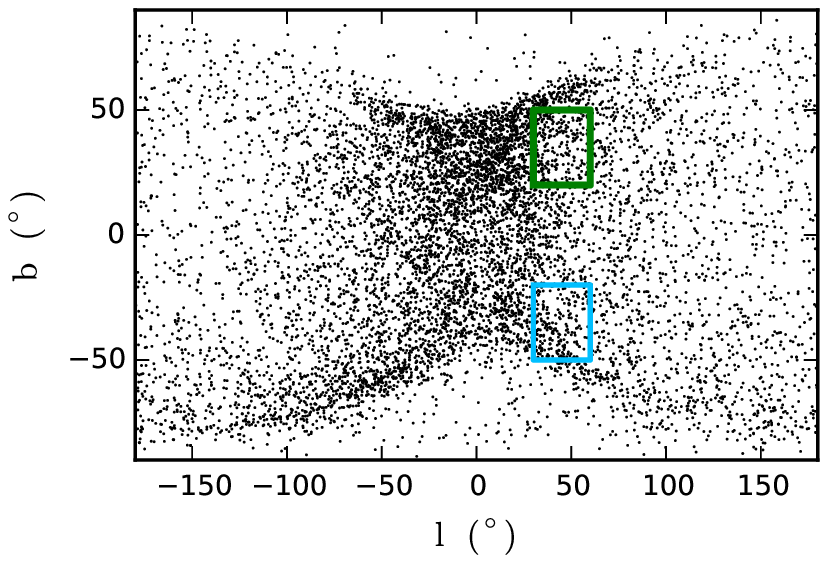}
\includegraphics[scale=0.95]{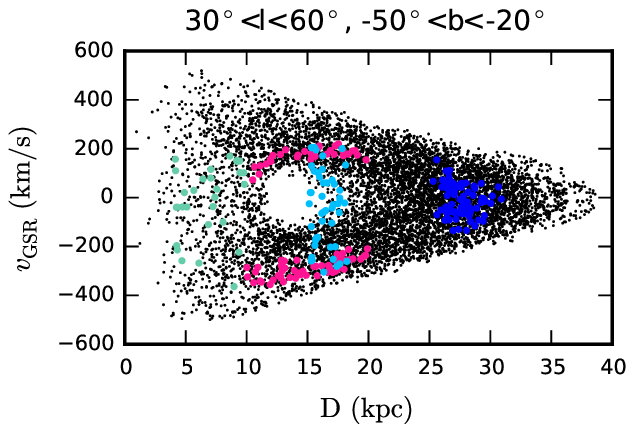}\\
\includegraphics[scale=0.9]{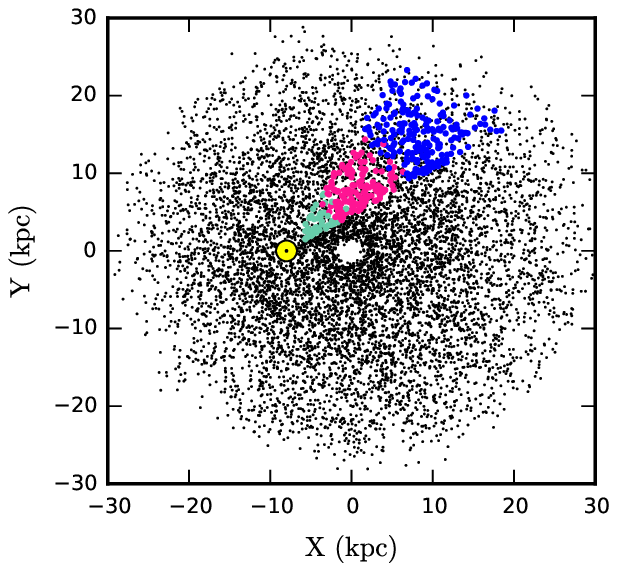}
\includegraphics[scale=0.9]{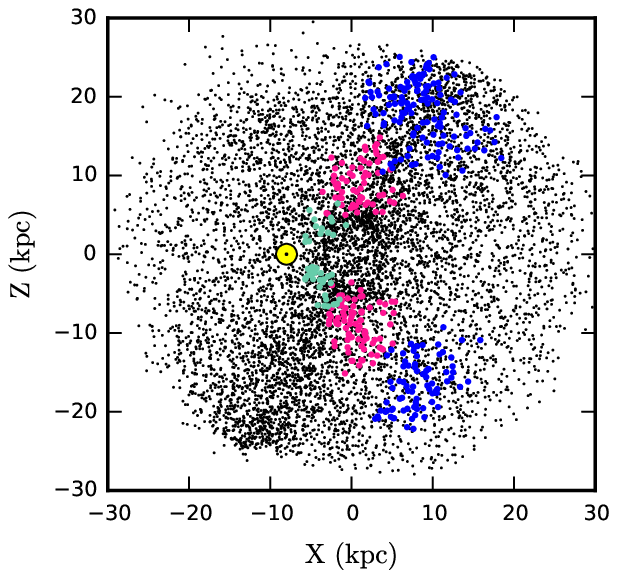}
\includegraphics[scale=0.9]{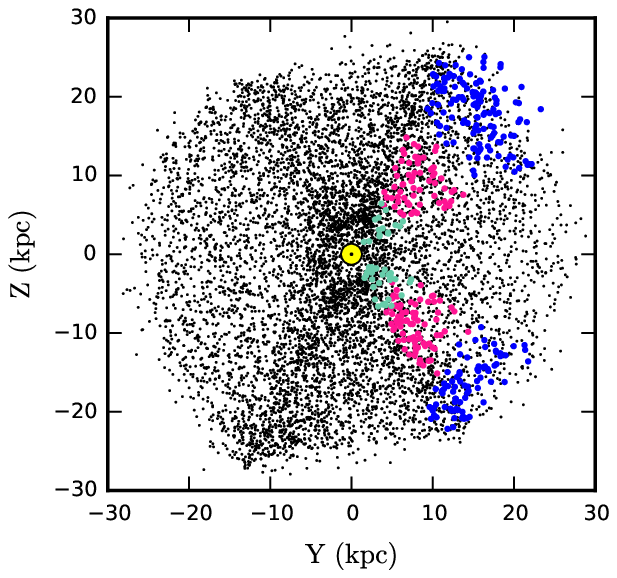}
\hspace{-1.0cm}
\includegraphics[scale=0.9]{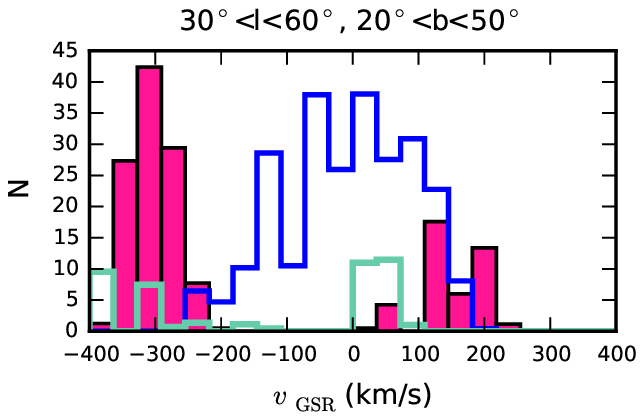}
\includegraphics[scale=0.9]{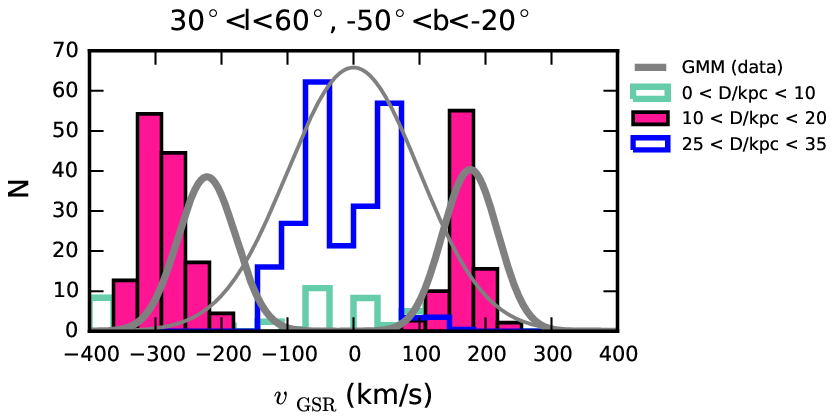}\\
\caption{Example of an N-body simulation of a disrupted satellite \citep{Bu05} orbiting in the inner regions of the halo a Milky Way type galaxy.  Top row: the debris in Galactic coordinates (left) and radial phase-space diagram for the simulated particles, from the Sun's perspective (right). Middle row: X-Y, X-Z and Y-Z projections of the particles' positions with the Sun's location marked in yellow. The observed RRL are overlaid over the simulation particles, in blue. The colours indicate three different selections along the debris: stars at distances with respect to the Galactic center similar to HAC are shown in pink ($20^{\circ} <l< 55^{\circ}$, $-45^{\circ}<b<-20^{\circ}$ and $15<D$/kpc$< 20 $); stars at pericenter in green  ($D<10$ kpc)  and stars at the apocenter in blue ($25<D/$kpc$<35$). Bottom panels: Mass-weighted velocity distribution of the particles in these three regions, in two symmetric fields (marked in cyan and green in the $l$-$b$ plane), corresponding to the Northern HAC (left panel) and the Southern HAC (right). The best fit GMM for all 4 tracers, shown in Fig.5, is marked in gray. }
\label{Bullock}
\end{figure*}

In this section we assess the possibility that the HAC structure represents debris from a disrupted MW satellite.

We visually inspect a suite of 11 stellar halo models built entirely from accretion events within the context of a $\Lambda$CDM universe \citep{Bu05}, to find examples of mixed debris in the inner halo ($r < 30$ kpc) that look similar to the HAC. We investigate the $(l,b)$ distributions of more than 1000 accretion events, seen from an internal perspective i.e., an Aitoff projection of the material as viewed from the Sun. As an example, we select a spatially well-mixed satellite (object number 100 from the simulation 'halo8', see Figure \ref{Bullock}) with a luminosity of $L \sim 10^{4} L_{\odot}$, accreted 11 Gyrs ago. The X-Y, X-Z and Y-Z projections of this satellite (middle panels of Figure~\ref{Bullock}) show the debris has a well-mixed morphology. It is less luminous and probably older than the accretion event that has generated the HAC, which has an estimated luminosity of  $10^{5}-10^{7} L_{\odot}$. 

The upper panels of Figure~\ref{Bullock} project the simulated particles in observable co-ordinates, as viewed from the assumed location of the Sun - at (X,Y,Z)=(-8,0,0) kpc, marked by the yellow symbol in the middle panels. The $l$-$b$ plane shows that, from the perspective of the Sun, which sits in the middle of the debris structure, the distinct edges of HAC might be interpreted as caustics (i.e. at $(l,b)\sim (0, \pm 50^\circ)$) marking the inner boundaries of a much larger density distribution structure that entirely encompasses the disc. The upper right panel of the figure shows the projection of all the simulated particles in line-of-sight velocity and distance from the perspective of the Sun. More realistically, spectroscopic observations are usually limited to a small patch of the sky and to a range of magnitudes therefore they can not map the full extent of the debris. In order to illustrate some possible perspectives on this structure, the coloured points in all panels mark three selections that have been chosen to show the variety of expected velocity distributions that might be seen in a limited area survey. We selected two symmetric fields which roughly correspond to the region where the HAC lies, $20^{\circ} <l< 55^{\circ}$ and $25^{\circ}<b<45^{\circ}$ (green box in the $l-b$ plane) and $20^{\circ} <l< 55^{\circ}$, $-45^{\circ}<b<-20^{\circ}$ (blue box), in three Heliocentric distance ranges: 
\begin {itemize}
\item with pink we mark the stars at intermediate distances, $15<D$/kpc$< 20$, where \citet{Si14} found the largest excess of RRL. The mass-weighted velocity distribution of these stars in the Southern field, shown in the bottom right panel of Figure \ref{Bullock}, exhibits two strong peaks: one at positive velocities $<v> \sim 180$ km/s and one at negative velocities $<v> \sim -310$ km/s in agreement with the double peaked distribution in RVs of the stellar tracers (the best fit GMM for all 4 populations is shown in gray). The mass-weighted velocity distribution of the stars in the symmetric field in the Northern Hemisphere (bottom left panel of Figure 7) has a strong peak at negative velocities and a peak at ~180 km/s which is in agreement with the red giant branch (RGB) stars velocities \citep{Be07}. It is perhaps not surprising that the velocity distribution of the RGBs in the north is single-peaked instead of double-peaked, if compared to what is expected from this simulation, which in the North is strongly dominated by one velocity peak.
\item  with blue we mark the points close to the apocenter, at $25<D/$kpc$<35$. Here we have the highest density of stars because the debris is most likely to be found at the orbital apocenter of the parent satellite, where the stars spend most of their time.  The line-of-sight velocities are broadly distributed around 0 km/s and the kinematic signal of the satellite could be confused to the background halo population ($\mu_{\mathrm{halo}} = 0$ km/s).
 \item with green we mark the points close to the pericenter, at $D < 10$ kpc. These stars, closer to the Sun, exhibit a uniform velocity distribution.
 \end{itemize}

We conclude that observations of HAC, both in morphology and in the double peaked velocity distributions apparent along with the expected halo field population seen in our samples are broadly consistent with the typical properties of debris from a satellite fully-mixed throughout the inner Galaxy. 
Studies of RR Lyrae probing different regions of this structure could map out how the velocity distribution varies across the sky and hence allow a more specific interpretation in the future.

\section{Results and Conclusions}
We designed a follow-up program to map the velocity distribution of the HAC, crucial for testing the dynamical predictions of stellar clouds. 
We measured the radial velocities of 45 RR Lyrae in the  southern portion of the HAC using ModSpec on the 2.4m Hiltner telescope at MDM. The observations took place during 6 nights between the 29th of August 2014 and the 3rd of September 2014. 

The basic data reduction steps (bias subtraction, flat fielding, wavelength calibration, extraction) were performed using IRAF routines. We have produced a spectral fitting code which we used to successfully model all the co-added RR Lyrae spectra and find the heliocentric radial velocities of the stars. These velocities were then corrected for variations due to stellar pulsations, producing the first large sample of velocities in the HAC. 

We have performed a multi-Gaussian decomposition of the velocity distribution in the HAC region and found it is best described by three Gaussians. The kinematic information from other tracers is not in disagreement with our findings with RR Lyrae: in addition to a halo population modeled with a Gaussian centred at 0 km/s and $\sigma = 105$ km/s, two other Gaussian components are required at moderately large negative and positive radial velocities ($\approx -200$ km/s and $\approx +200$ km/s).

To provide a possible interpretation for our results, Re have used a suite of N-body simulations from \citet{Bu05}. The behaviour uncovered in the RR Lyrae sample (and supported, at least in part, by other tracers) is typical of an old accretion event with small apo-galactic radius. In these events, at redshift $z=0$, the debris is fully wrapped up in phase-space with familiar "chevron" features overlapping. 
If we are observing the HAC along the debris (for example where the stars marked in pink in Figure \ref{Bullock} are situated), then the velocity distribution is bound to show an excess at moderate negative and positive velocities, as in the data. However, if we select particles at the apocenters, the velocity distribution would only show a broad component centred on $V = 0$ km/s, while close to the pericenter the distribution would be almost flat. Mapping the velocity distribution of associated RR Lyrae across the sky would allow a more definitive interpretation of this interesting structure. 

\section*{Acknowledgements}
The research leading to these results has received funding from the European Research Council under the European Union’s Seventh Framework Programme (FP/2007-2013) through the Gaia Research for European Astronomy Training (GREAT-ITN) Marie Curie Network Grant Agreement n. 264895 and the ERC Grant Agreement n. 308024. SK thanks the United Kingdom Science and Technology Council (STFC) for the award of Ernest Rutherford fellowship (grant number ST/N004493/1). KVJ's contributions were supported by NSF grant AST-1614743. The authors would like to thank the anonymous referee for their useful comments.

\bibliographystyle{mn2e}
\bibliography{bibl}  
\end{document}